\documentclass[%
reprint,
superscriptaddress,
 amsmath,amssymb,
 aps,
]{revtex4-2}

\usepackage{graphicx}
\usepackage{dcolumn}
\usepackage{bm}
\usepackage[version=3]{mhchem} 
\usepackage{booktabs} 
\usepackage{hyperref}
\hypersetup{colorlinks=true, linkcolor=blue, citecolor=blue, urlcolor=blue} 

\newcommand{\topic}[1]{}

\begin{document}

\preprint{APS}

\title{Engineering Interfacial Charge Transfer through Modulation Doping for 2D Electronics}

\author{Raagya Arora}
\email{raagya@g.harvard.edu}
\affiliation{John A. Paulson School of Engineering
and Applied Sciences, Harvard
University, Cambridge, Massachusetts 02138, United States}

\author{Ariel R. Barr}
\affiliation
{Department of Materials Science \& Engineering, Massachusetts Institute of Technology, Cambridge, Massachusetts 02139, United States}
\author{Daniel T. Larson}
\affiliation{Department of Physics, Harvard
University, Cambridge, Massachusetts 02138, United States}
\author{Michele Pizzochero}
\affiliation{Department of Physics, University of Bath, Bath BA2 7AY, United Kingdom}
\affiliation{John A. Paulson School of Engineering and Applied Sciences, Harvard University, Cambridge, Massachusetts 02138, United States}
\author{Efthimios Kaxiras}
\email{kaxiras@g.harvard.edu}
\affiliation{Department of Physics, Harvard
University, Cambridge, Massachusetts 02138, United States}
\affiliation{John A. Paulson School of Engineering
and Applied Sciences, Harvard
University, Cambridge, Massachusetts 02138, United States}


\date{\today}

\begin{abstract}
  

Two-dimensional (2D) semiconductors are likely to dominate next-generation electronics due to their advantages in compactness and low power consumption. 
However, challenges such as high contact resistance and inefficient doping hinder their applicability. 
Here, we investigate work-function-mediated charge transfer (modulation doping) as a pathway for achieving high-performance p-type 2D transistors. 
Focusing on type-III band alignment, we explore the doping capabilities of 27 candidate materials, including transition metal oxides, oxyhalides, and $\alpha$-RuCl$_3$, on channel materials such as transition metal dichalcogenides (TMDs) and group-III nitrides.  Our 
extensive first-principles density functional theory (DFT) reveal p-type doping capabilities of high electron affinity materials, including $\alpha$-RuCl$_3$, MoO$_3$, and V$_2$O$_5$ .  We predict significant reductions in contact resistance and enhanced channel mobility through efficient hole transfer without introducing detrimental defects.  We analyze transistor geometries and identify promising material combinations beyond the current focus on WSe$_2$ doping, suggesting new avenues for hBN, AlN, GaN, and MoS$_2$. This comprehensive investigation provides a roadmap for developing high-performance p-type monolayer transistors toward the realization of 2D electronics.
\end{abstract}

\maketitle

\section{Introduction}

\topic{2D materials are important}
Layered two-dimensional (2D) semiconductors have attracted  attention for their potential to revolutionize electronic devices through miniaturization. The 2D materials family encompasses diverse behavior, including semi-metallic graphene, semiconducting transition metal dichalcogenides (TMDs) and black phosphorus (BP), and insulating hexagonal boron nitride (hBN)~\cite{chaves2020bandgap, novoselov20162d}. These materials exhibit exceptional electronic and optical properties, making them highly promising candidates for applications in field-effect transistors (FETs), photovoltaic cells, photodetectors, spintronics, sensors, and flexible electronics~\cite{xie2024low, wu2024synthesis, akinwande2019graphene, radisavljevic2011single, sangwan2018multi}.

\topic{Challenges: contact resistance and carrier concentration}
Despite extensive research demonstrating the superior performance of 2D material-based nanodevices over bulk semiconductors, significant challenges still need to be overcome to achieve practical technological advancements. Two crucial aspects relevant to our study are:
(i)  the electrical contact between semiconductors and metal electrodes, which governs the charge carrier injection from metal to the 2D channel;  and 
(ii) the doping of the semiconductor layers, which controls the carrier concentration~\cite{akinwande2019graphene}.

\topic{contact resistance}
The first issue, high contact resistance, hampers circuit operation, leading to slower performance, unnecessary heat generation, a reduced on-off ratio, and increased power consumption.
When a semiconductor is heavily doped near the metal contact, the increased carrier concentration reduces the width of the depletion region at the interface. This effectively thins the Schottky barrier, facilitating carrier transport through enhanced tunneling. 
Consequently, contact resistance is lowered, and control over the electrical conduction within the channel layer is improved.  
However, due to their atomically thin nature, 2D materials do not favor substitutional doping techniques such as ion implantation, which involves bombardment with impurities, because it tends to reduce carrier mobility. 
Progress in alternative doping techniques, such as electrostatic gating, is still limited by issues like electrical breakdown or the need for sophisticated device fabrication~\cite{wu2023electrostatic}.
While advancement of n-type contacts is progressing, research into reliable and low-resistance p-type contacts is somewhat lagging.\cite{wang2022p, patoary2023improvements, mcclellan2021high}

\topic{carrier concentration}
In what concerns the second issue, adjusting the type and concentration of charge carriers to manipulate electrical conduction has been a fundamental capability driving the progress of electronics and optoelectronics~\cite{seo2021reconfigurable,wang2021modulation}. 
FETs made from 2D materials generally suffer from reduced carrier mobility, which has been recognized as a major limiting factor in harnessing the full potential of these materials for next-generation nanoelectronics. This limitation can be due to factors like scattering from defects, impurities, or interactions with the substrate the 2D material is placed on.
To address both limitations of contact resistance and low carrier concentration, various doping techniques have been explored~\cite{seo2021reconfigurable,wang2021modulation,ho2023hysteresis}.


 
\topic{Doping method considered here}
Maintaining the structural integrity of 2D materials during doping is crucial, as defects can negatively affect the carrier mobility. Effective doping should enhance both carrier density and mobility. In the present study we propose several options for layer combinations that can effectively p-dope 2D channel materials through work-function-mediated interlayer charge transfer~\cite{shao2021work}. This method, also referred to as ``modulation doping'' or ``remote charge transfer doping'', involves providing additional carriers through a separate proximal layer~\cite{lee2021remote, wang2020modulation, ozccelik2016band}. 
Instead of introducing defects into the 2D channel material, the channel forms a heterostructure with materials possessing extreme work functions, favoring band alignment such that the layer donates electrons or holes to the channel layer, thereby allowing modulation of the channel material's conductivity.   This strategy is also called surface transfer doping (STD) since doping is achieved by electron exchange at the interface between the channel surface and a solid dopant~\cite{ristein2006surface}.
This method bypasses surface treatments and metal evaporation, simplifying p-type doping in 2D semiconductors.

Examples of transistor geometries that favor reductions in contact resistance and/or improved channel mobility, high carrier density are shown in Fig.~\ref{fig:Figure1}.
Using first-principles calculations, we assess the hole transfer capabilities of transition metal oxide films and $\alpha$-RuCl$_3$, demonstrating robust p-type doping of the channel material, with potential for improving both contact resistance and carrier concentration.
We present a list of favorable
layer combinations that exhibit charge transfer and emphasize the electron acceptor properties underlining their capacity to reduce contact resistance and enhance channel mobility.


\section{RESULTS AND DISCUSSION}

\topic{specific channel materials and their challenges}
We envision devices where the 2D channel is a hexagonal group III nitride or TMD. 
Hexagonal 2D materials such as hBN and its analogs (hAlN, hGaN) offer exceptional electronic, optical, and thermoelectric properties, along with high mechanical strength. 
Achieving reliable n-type or p-type doping in these materials has proven to be challenging~\cite{lu2022towards}. Similarly, TMDs have garnered significant attention due to their exceptional properties, including high carrier mobilities, large bandgaps, and atomic thickness~\cite{wang2024critical}. Creating electrical contacts to monolayer TMDs is hindered by the common challenges associated with semiconductor contacts, including Schottky barrier formation and Fermi-level pinning\cite{wang2022p}.

\begin{figure*}
        \centering
        \includegraphics[width=1\linewidth]{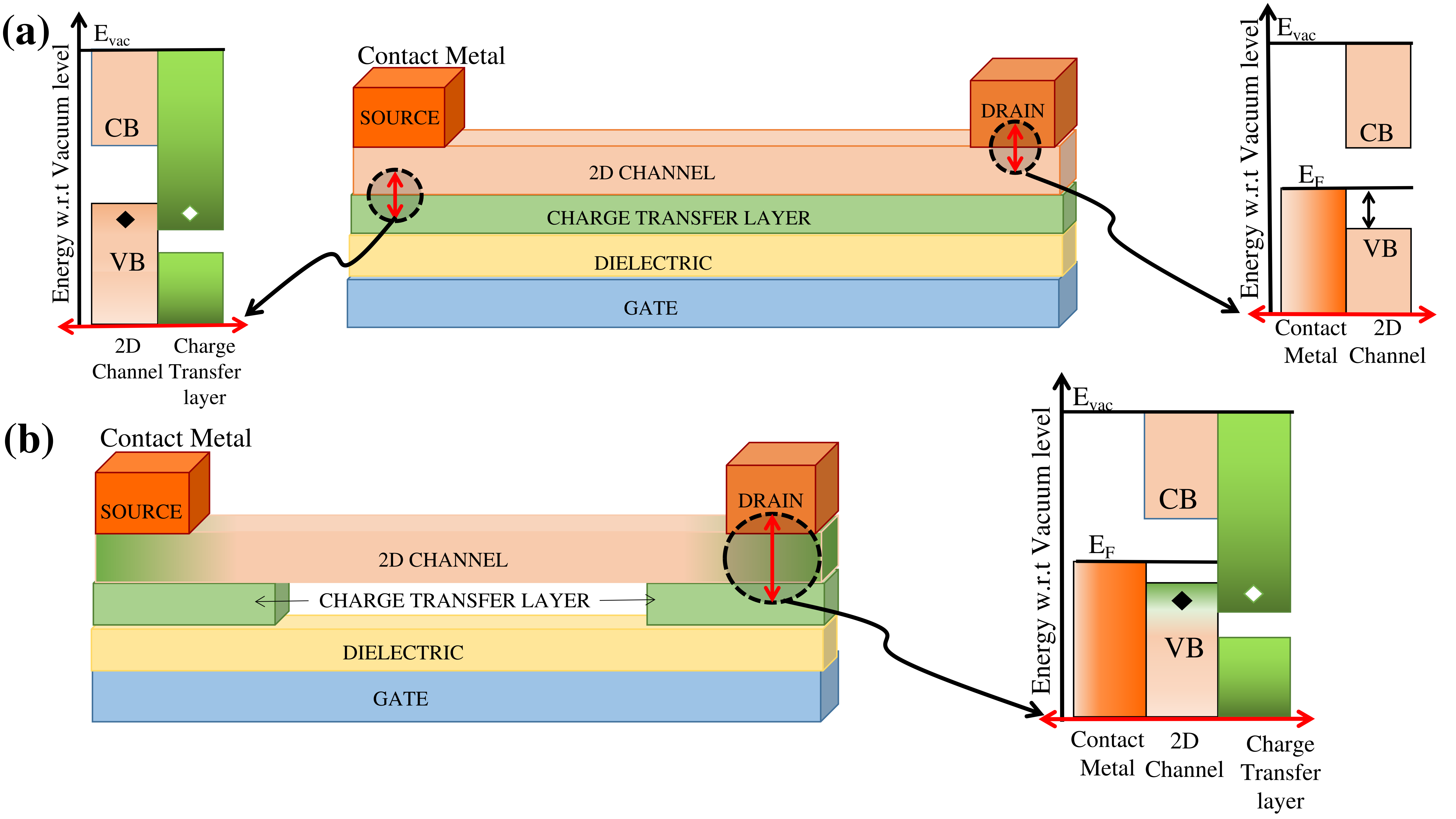}
        \caption{Schematic device structure and electronic bands for modulation doping:   (a) Source, drain, channel, and charge transfer layer (CTL) of a 2D-channel-based transistor. The band diagrams on each side illustrate the separate interfaces channel/CTL and drain/2D-channel. The difference in work function between the channel and CTL modifies the electronic bands, resulting in the charge transfer layer acting as a so-called ‘van der Waals’ electron acceptor, while the channel becomes significantly hole-doped after transferring its electrons. (b) Device with CTL localized near the contact, creating regions in the channel that are highly hole-doped. The band diagram on the right illustrates the expected charge redistribution when the channel and charge transfer layers are brought into contact. The modification of the channel work function serves to reduce the Schottky barrier.}
        \label{fig:Figure1} 
    \end{figure*}

\topic{Charge tranfer layers}
In order to dope those 2D channels, materials with high work functions like $\alpha$-RuCl$_3$ have aroused interest for their capacity to function as a charge transfer layer in FET geometries like those schematically shown in Fig.~\ref{fig:Figure1}. Studies have explored surface charge-transfer doping of 2D WSe$_2$, achieving superior electrical performance, including a hole density of $4 \times 10^{13}$ cm$^{-2}$ and a hole mobility exceeding 200 cm$^2$/Vs~\cite{xie2024low}. 
Such levels of doping are challenging to achieve with conventional electrostatic gating due to dielectric breakdown.
Oxide materials such as MoO$_3$ and V$_2$O$_5$ are also promising candidates for the STD process~\cite{crawford2018thermally, kuruvila2014organic}. Notably, MoO$_3$'s exceptionally high work function has led to its widespread adoption as an effective anode buffer layer, significantly enhancing hole injection and extraction in organic electronic devices~\cite{butler2015band, meyer2014metal}. Moreover, MoO$_3$ functions as a bulk p-type dopant for organic wide-band-gap materials~\cite{greiner2012universal, greiner2013thin}.

\topic{Definition of IP and EA}
The ionization potential (IP) and electron affinity (EA) of a semiconductor or insulator provide crucial insights into their electronic behavior, indicating the ease with which an electron can be released or accepted, respectively. These properties are defined by the energy levels of the valence band maximum (VBM) and conduction band minimum (CBM) relative to the vacuum level ($E_{\rm vac}$), namely: $E_\mathrm{IP} = E_\mathrm{\rm vac} - E_\mathrm{VBM}$ and $E_\mathrm{EA} = E_\mathrm{\rm vac} - E_\mathrm{CBM}$. Understanding the relative positions of these energy levels across different materials, known as band alignment, is of central importance for the design of devices that effectively utilize interfacial and surface-related phenomena.

\topic{Type-III band alignment}
We focus on heterojunctions with type-III band alignment, also known as a broken gap,  characterized by a substantial energy offset such that the CBM of one material lies below the VBM of an adjacent material, resulting in a direct overlap between those two band manifolds. 
This broken-gap configuration induces spontaneous charge transfer. The applicability of type-III band alignment in advanced devices is particularly beneficial for tunneling-based applications, such as tunneling field-effect transistors, infrared detectors, and photodetectors, where low-power operation and efficient charge transfer are essential \cite{lei2019broken,yang2020epitaxial, zhao2023curvature,guo2024composition}.
The heterostructure formed from WSe$_2$ and $\alpha$-RuCl$_3$ offers a concrete example of how type-III band alignment works: the CBM of $\alpha$-RuCl$_3$ is positioned below the VBM of WSe$_2$, facilitating spontaneous charge transfer from WSe$_2$ to $\alpha$-RuCl$_3$ and resulting in a heavily p-doped WSe$_2$ layer\cite{pack2024charge,xie2024low}.

\topic{Summary of computational method}
We employed first-principles density functional theory (DFT) calculations using the HSE06 functional to determine $E_\mathrm{IP}$ and $E_\mathrm{EA}$ for 21 binary oxides, 2 MXO (M: Metal, X: Halide, O: Oxygen), 3 A$_2$BO$_4$ layers (A: Ba/Ca/Sr, B: Ti, O: Oxygen), and $\alpha$-RuCl$_3$. This computational approach allows us to rigorously analyze the band alignment of these materials and identify optimal candidates for effective p-type doping of TMDs and group III nitride materials (for details see Supplementary Information, SI). The essential criterion for selecting promising materials is that they must possess a sufficiently high work function for hole doping or a sufficiently low work function for electron doping. Devices built from 2D materials with appropriately aligned energy levels will have enhanced performance due to improved charge carrier injection and reduced contact resistance.

\topic{Modulation doping layer, Fig.1a}
Incorporating a modulation doping layer can enhance two critical properties of FETs. 
Fig.~\ref{fig:Figure1}(a) shows the charge-transfer architecture, wherein a van der Waals (vdW) electron acceptor material, such as those studied here, is placed in contact with the entire 2D channel. 
This approach alters the doping state of the 2D channel, resulting in higher carrier mobility compared to intrinsically doped TMD-based transistors\cite{cho2022modulation}. 
Modifications to this geometry, incorporating a vdW hBN spacer layer between the charge transfer layer and the 2D channel, offer a compelling route to enhance device performance by enabling effective confinement of doped carriers. Recent experimental strategies have successfully demonstrated the efficacy of inserting vdW hBN spacer layers in similar device architectures\cite{lee2021remote}.

\topic{Modulation doping layer, Fig.1b}
Fig.~\ref{fig:Figure1}(b) presents an alternative device architecture incorporating charge transfer layers to selectively dope the contact regions.  Within this structure, the contact metal largely dictates the Fermi level within the 2D channel. Consequently, the electron Schottky barrier height (SBH) corresponds to the energy difference between the Fermi level and the CBM, while the hole SBH is defined as the energy separation between the Fermi level and the VBM, as illustrated in the right panel of Fig.~\ref{fig:Figure1}(a). By modifying the modulation doping layer, we can effectively tune the Fermi level position across a significant portion of the 2D semiconductor's band gap. 
This tunability allows for a reduction in the SBH for either hole or electron doping, as shown in the right panel of Fig.~\ref{fig:Figure1}(b), highlighting the potential for optimizing device performance through modulation doping\cite{cho2022modulation}. 
This strategy aligns with recent work by Jordan Pack {\em et al.}\cite{pack2024charge}, who demonstrated that $\alpha$-RuCl$_3$ can exclusively dope the contact region of an hBN-encapsulated monolayer WSe$_2$ device.  While the experimental availability of these materials makes the creation of such heterostructures attainable, their (electro-)chemical reactivity necessitates meticulous optimization.

\begin{figure*}
    \centering
    \includegraphics[width=0.8\linewidth]{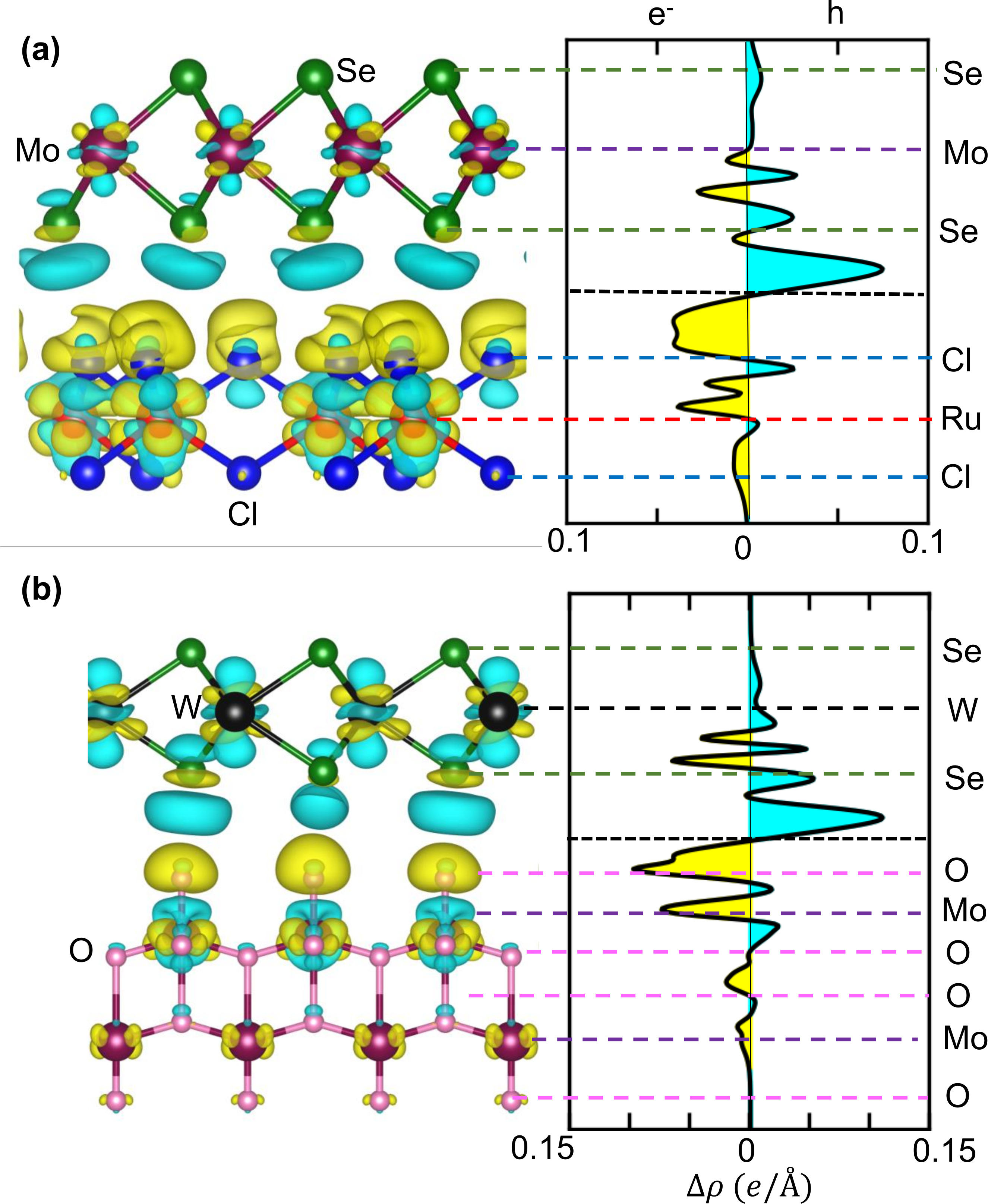}
    \caption{Structures and isosurfaces of charge density difference between the combined hetrostructure and its constituent layers, for: (a) MoSe$_2$ and $\alpha$-RuCl$_3$; (b) WSe$_2$ and  MoO$_3$.
    Yellow contours show regions of electronic charge accumulation (extra electrons) and cyan contours indicate charge depletion (holes), demonstrating an overall transfer of electrons from the 2D channel to the charge transfer layer.
    Plots show the charge density difference averaged over horizontal planes as a function of the perpendicular (vertical) distance.  }
    \label{fig:Figure2}
\end{figure*}

\topic{Microscopic view of charge transfer}
To understand better the microscopic details of charge transfer, we calculated the charge density difference for specific pairings of channel materials and doping layers. The charge density difference, $\Delta \rho$, is defined as:
\begin{equation}
\Delta \rho = \rho_{\text{A/B}} - \rho_{\text{A}} - \rho_{\text{B}} 
\end{equation}
where $\rho_\mathrm{A/B}$, $\rho_\mathrm{A}$, and $\rho_\mathrm{B}$ denote the charge densities of the combined heterostructure, the isolated layer A, and the isolated layer B, respectively.
We show two representative channel material/doping layer pairs in Fig.~\ref{fig:Figure2}: (a) WSe$_2$ with $\alpha$-RuCl$_3$, and (b) MoSe$_2$ with $\alpha$-MoO$_3$. To elucidate the charge rearrangement and interlayer coupling mechanism 
in these heterostructures, 
we calculated the plane-averaged charge density difference along the $z$-axis:  
yellow isosurfaces of the charge density difference indicate charge accumulation (excess electrons), while cyan isosurfaces indicate charge depletion (excess holes). As is clear from the results shown in Fig.~\ref{fig:Figure2}, charge depletion predominantly occurs within the WSe$_2$ and MoSe$_2$ layers, accompanied by a corresponding accumulation of charge on $\alpha$-RuCl$_3$ and MoO$_3$. 
Other heterostructures, for example, $\alpha$-RuCl$_3$/WSe$_2$ and MoO$_3$/MoSe$_2$, exhibit qualitatively similar features (see SI).

\begin{figure*}
    \centering
    \includegraphics[width=0.9\linewidth]{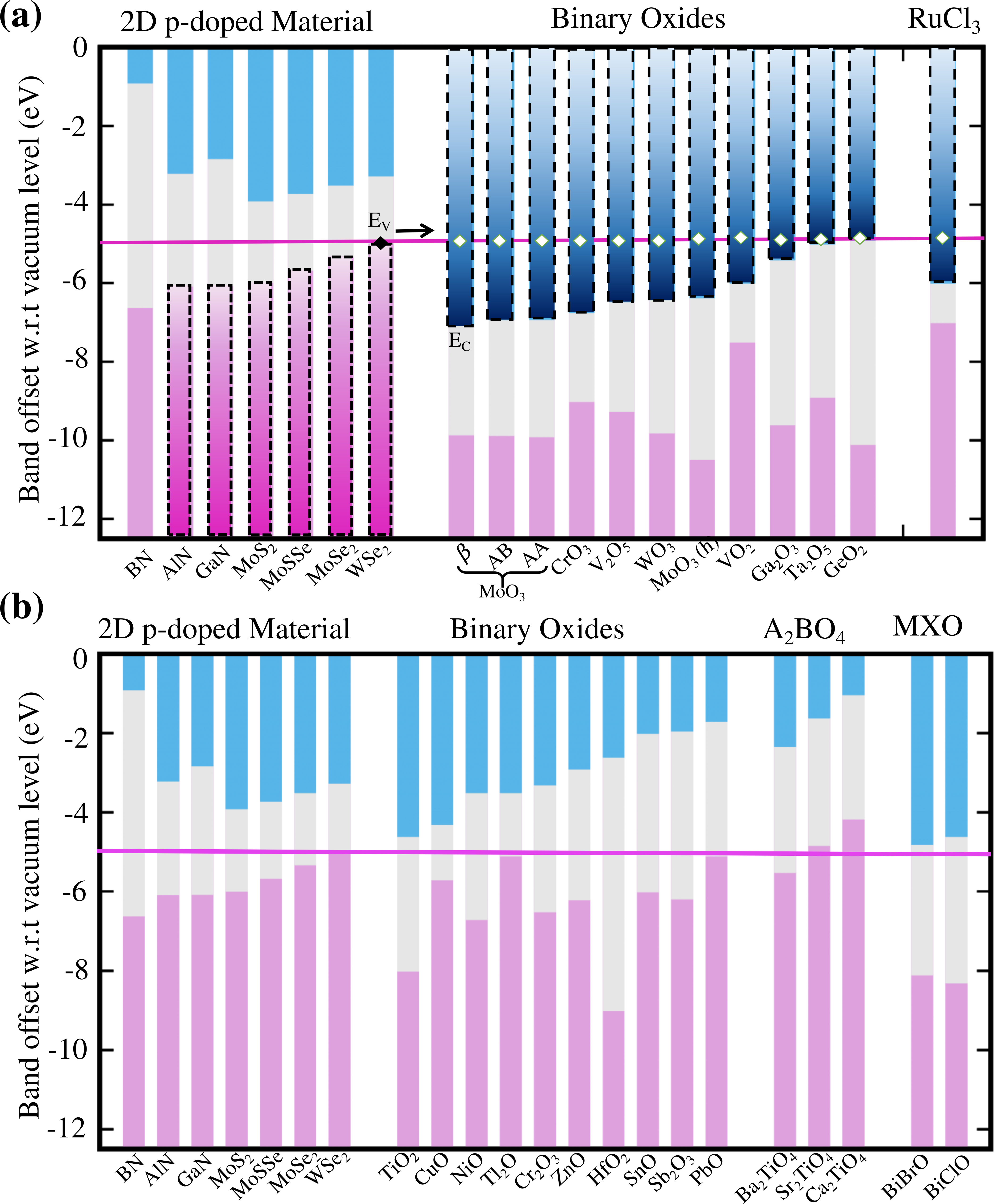}
    \caption{(a) Band Alignments for p-Type Doping in 2D Materials. Valence and conduction band positions are shown with respect to the vacuum level for various 2D channel materials (left) and binary oxides acting as acceptor layers. The horizontal pink line, aligned with the VBM of WSe$_2$, highlights favorable type-III band alignments for p-type doping.  Solid diamonds and black dashed lines indicate spontaneous charge transfer between WSe$_2$ and the suitable oxides.  $\alpha$-RuCl$_3$ also acts as an effective electron acceptor for all channel materials but hBN. (b) Band Alignments Lacking Type-III Character. VBM and CBM positions, referenced to the vacuum level, for binary oxides, A$_{2}$BO$_4$ pervoskite monolayers and  MXOxides (M: metal, X: halide) that do not exhibit a type-III band alignment with the considered channel materials.
    }
    \label{fig:Figure3}
\end{figure*}

\topic{Band alignments Fig, top}
In Fig.~\ref{fig:Figure3} we provide the theoretical ionization potentials ($E_\mathrm{IP}$) and electron affinities ($E_\mathrm{EA}$) for a selection of different channel materials and doping layers. We considered seven 2D channel materials, all experimentally relevant for 2D semiconductor FETs, namely the
transition-metal dichalcogenides, MoS\textsubscript{2}, WSe\textsubscript{2}, MoSe\textsubscript{2}, and MoSSe, as well as 2D hexagonal AlN, BN, and GaN. 
These 2D channel materials are arranged in ascending order of the VBM relative to the vacuum level (corresponding to decreasing $E_\mathrm{IP}$). WSe\textsubscript{2} exhibits the lowest ionization potential, $E_\mathrm{IP}= 4.99$ eV, making it the most suitable candidate for type-III band alignment. 
Conversely, h-BN has its VBM situated far from the vacuum level, $E_\mathrm{IP}=6.61$ eV, which complicates the search for appropriate dopants. 

\topic{Discussion of specific modulation doping materials}
We considered a wide range of 2D or layered materials as potential charge transfer layers, including many binary transition metal oxides, as well as $\alpha$-RuCl$_3$, oxyhalides, and perovskite layers. 
Fig.~\ref{fig:Figure3}(a) shows the successful combinations of channel and doping layers for which the theoretically predicted band alignment makes $p$-doping of WSe$_2$ possible. 
In Fig.~\ref{fig:Figure3}(b) we also show the {\em unsuccessful} combinations of materials, where the doping layer does \emph{not} exhibit a type-III band alignment with any of the investigated 2D channels. 
Below we provide a summary of the various charge transfer layers and highlight relevant materials based on their IPs and EAs.

We begin with $\alpha$-RuCl\textsubscript{3}, which is known to significantly hole-dope graphene and WSe$_2$~\cite{wang2020modulation,xie2024low}. This narrow-band Mott insulator exhibits a notably deep work function, $E_\mathrm{EA} = 6.1$ eV (see Fig.~\ref{fig:Figure3}), which significantly surpasses the typical work functions of common layered materials such as graphene ($E_\mathrm{IP}=4.6$ eV) and WSe\textsubscript{2} ($E_\mathrm{IP}=4.4$ eV). The CBM of  
$\alpha$-RuCl\textsubscript{3} is far away from the vacuum level, suggesting its potential applicability as an electron acceptor not only for graphene and WSe\textsubscript{2}, but potentially for a wider range of 2D materials, particularly all of the 2D TMD channel materials and group III nitrides we considered except for BN.
This combination of a deep work function and CBM distinguishes $\alpha$-RuCl\textsubscript{3} as a important material within this context.

High electron affinity transition metal oxides, such as molybdenum trioxide (AA MoO\textsubscript{3}, $E_\mathrm{EA} = 6.4$ eV), tungsten trioxide (WO\textsubscript{3}, $E_\mathrm{EA} = 6.4$ eV), and vanadium pentoxide (V\textsubscript{2}O\textsubscript{5}, $E_\mathrm{EA} = 6.46$ eV), are well-established and highly effective surface acceptors\cite{yun2017aligning,hu2023vanadium}. MoO\textsubscript{3} exhibits various phases that have been studied under different synthesis conditions. These phases include the thermodynamically stable and most well studied $\alpha$-MoO\textsubscript{3} (AA) and several metastable forms such as II-MoO\textsubscript{3} (AB), h-MoO\textsubscript{3}, and $\beta$-MoO\textsubscript{3}. The fundamental structure of MoO\textsubscript{3} is based on MoO\textsubscript{6} octahedra, where a central molybdenum atom is surrounded by six oxygen atoms.
The electronic structure of bulk MoO\textsubscript{3} surfaces and interfaces with organic semiconductors has been studied theoretically using DFT calculations, but its effect on the energy band levels of diverse 2D channel materials for FET applications remains an open question. 
Our study highlights stable $\alpha$-MoO\textsubscript{3}, as well as metastable II-MoO\textsubscript{3}, h-MoO\textsubscript{3}, and $\beta$-MoO\textsubscript{3}, as promising candidates for modulating the electronic properties of all the 2D channel materials we investigated. While current FET literature primarily focuses on doping WSe\textsubscript{2} with MoO\textsubscript{3}, our findings suggest exciting new avenues for doping other materials, including hBN, AlN, GaN, and MoS\textsubscript{2} — a prospect that warrants further investigation.
$\alpha$-MoO$_3$ facilitates spontaneous electron transfer from WSe$_2$ to itself, effectively injecting holes into the WSe$_2$ layer in agreement with the band alignment presented in Fig.~\ref{fig:Figure3}\cite{cai2017rapid}.
WO\textsubscript{3} shares the same basic structure with MoO\textsubscript{3} and has an electron affinity of $6.4$ eV, making it also a promising candidate for type-III band alignment (see Fig.~\ref{fig:Figure3}).\cite{wang2011electronic}

The prevalent $\alpha$ phase of V\textsubscript{2}O\textsubscript{5}, characterized by the Pmnm space group, features a structure where single layers consist of edge- and corner-sharing square pyramids, with adjacent layers bound together along the c-axis through weak vdW interactions. Notably, V\textsubscript{2}O\textsubscript{5} boasts a layered structure, a direct band gap within the visible-light region, and impressive chemical and thermal stability. 
This material is easy to prepare and has been used in heterostructure engineering with graphene and MXenes to enhance cathode performance\cite{falun2024enhancing}. V\textsubscript{2}O\textsubscript{5} has an extremely high electron affinity (6.46 eV), which makes it suitable for hole doping in six of the seven 2D materials evaluated in this study (see Fig.~\ref{fig:Figure3}).

The crystal structure of chromium trioxide (CrO\textsubscript{3}) consists of chromium atoms surrounded by a tetrahedron of four oxygen atoms. The corner-sharing tetrahedra form a 
chain-like arrangement, wherein each Cr atom shares two O atoms with its adjacent Cr atoms along the chain. CrO\textsubscript{3} has an extremely high electron affinity (6.75 eV) suitable for hole doping all seven 2D channel materials evaluated (see Fig.~\ref{fig:Figure3}). 

Layered Ga\textsubscript{2}O\textsubscript{3} exhibits to type-III band alignment in Ga\textsubscript{2}O\textsubscript{3}/SiC vdW heterostructures\cite{ferdous2024resilient}, and our calculations show that it will likely p-dope WSe$_2$ and possibly MoSe$_2$. We also predict p-type doping of WSe$_2$ using Ta\textsubscript{2}O\textsubscript{5}, which is stable in an orthorhombic phase. 
Monolayer GeO$_2$ warrants consideration as well. Its extremely wide excitonic gap of 6.24 eV~\cite{sozen2021vibrational}, positions it differently compared to the other materials investigated. This substantial bandgap, coupled with a relatively high work function ($E_\mathrm{EA}=5$ eV)  offer distinctive charge transfer characteristics. The successful fabrication of 2D GeO$_2$ further motivates its exploration\cite{zhang2021hexagonal}.

\topic{Combinations without Type-III alignment}
 
Even among the potential charge transfer layer combinations that do not exhibit type-III band alignment, shown in Fig.~\ref{fig:Figure3}(b), several are worth noting. 
Monolayers of the form A$_2$BO$_4$ are derived from cubic perovskites (ABO$_3$) and can be thought of as a single layer of the Ruddlesden-Popper phase with $n=1$~\cite{RP-phase-1958}.
Though they cannot serve as charge transfer layers for any of the considered 2D channel materials, since the energy of their valence bands is comparable to that of WSe$_2$ they could potentially be $p$-doped themselves by many of the binary oxides we studied.
The oxyhalides BiBrO and BiClO, which possess extremely high electron affinities~\cite{ran2018bismuth, prasad20202d}, 
are also interesting. Their bands are not far from type-III alignment with WSe$_2$ and MoSe$_2$, suggesting they may serve as potential p-type dopants, especially if defects or other surface details tip the balance in their favor.\\

\topic{Limitations of DFT for realistic interfaces}
\section{SUMMARY AND CONCLUSIONS}

The main results of this study are contained in 
Fig.~\ref{fig:Figure3}: part (a) showcases many promising candidates for p-type doping of 2D materials, all warranting further exploration and experimental validation; conversely, 
part (b) identifies candidate combinations that are not likely to work. 
Our systematic investigation of band alignments of 27 materials including oxides, oxyhalides, and $\alpha$-RuCl$_3$, with respect to TMDs and III-nitrides, provides a comprehensive foundation for future explorations of vdW heterostructures with optimized interfacial electronic properties for advanced electronic and optoelectronic devices.

Moreover, our study motivates a broader investigation into the doping capabilities of sub-stoichiometric metal oxides beyond the commonly explored MoO$_x$, AlO$_x$,
and TiO$_x$~\cite{shahrokhi20212d,mcclellan2021high}. The non-stoichiometry in these oxides can introduce vacancies and alter the electronic structure, showing great promise as dopants. Expanding the exploration of such materials could unlock new avenues for fine-tuning the electronic properties of 2D materials.

The electronic behavior of the interfaces between the materials studied here can be significantly influenced by a range of physical phenomena. These include the orientation and termination of the oxide surface, the presence of localized defects, and stress-induced distortions. Furthermore, calculations of band energies with respect to the vacuum level can be sensitive to the choice of exchange-correlation functional.  A comprehensive exploration of all these processes should be the subject of future work for more realistic predictions. 
\section*{Acknowledgement}
The authors thank Tomas Palacios,  Jing Kong, Jagadeesh Moodera and their research groups for valuable discussions. This work was supported mainly by Army Research Office Agreement Number W911NF-23-2-0057,
and partly by ARO Agreement Number W911NF-21-1-0184 and ARO Cooperative Agreement Number W911NF-21-2-0147. Calculations were performed on the FASRC cluster supported by the FAS Division of Science Research Computing Group at Harvard University.

\section*{Supplementary Information}
The Supplementary Information contains technical details and parameters of first-principles calculations along with additional results of the structural and electronic properties and related systems.


\appendix

\nocite{*}

\providecommand{\noopsort}[1]{}\providecommand{\singleletter}[1]{#1}%

\end{document}